\documentclass[a4paper,11pt]{article}
\pdfoutput=1
\usepackage[utf8]{inputenc}
\usepackage[T1]{fontenc}
\usepackage{jheppub}
\usepackage{natbib}

\usepackage[usenames,dvipsnames,svgnames,table,x11names]{xcolor}
\usepackage{lmodern,amsmath,multirow,cancel}
\usepackage{booktabs}
\usepackage{xstring}
\usepackage{ifthen}

\usepackage{ascmac,braket,bm,mathrsfs,amsthm,amsfonts}
\usepackage{hyperref}
\usepackage{graphicx}
\usepackage{subcaption} 
\usepackage{comment}
\usepackage[normalem]{ulem}

\usepackage{titlesec}
\newcommand*{\justifyheading}{\raggedright}
\titleformat{\chapter}[display]
  {\normalfont\huge\bfseries\justifyheading}{\chaptertitlename\ \thechapter}
  {20pt}{\Huge}
\titleformat{\section}
  {\normalfont\Large\bfseries\justifyheading}{\thesection}{1em}{}
\titleformat{\subsection}
  {\normalfont\large\bfseries\justifyheading}{\thesubsection}{1em}{}
\titleformat{\subsubsection}
  {\normalfont\bfseries\justifyheading}{\thesubsubsection}{1em}{}
\usepackage[english]{babel}
\usepackage[dvipsnames]{xcolor}
\numberwithin{equation}{section}

\usepackage{slashed}

\DeclareMathOperator{\Tr}{\mathrm{Tr}}

\renewcommand\Re{\mathop{\mathrm{Re}}}
\renewcommand\Im{\mathop{\mathrm{Im}}}

\usepackage{environ}
\newcommand{\rtext}[1]{\textcolor{red}{#1}}

\newcommand{\opn}[1]{\operatorname{#1}}
\NewEnviron{eqsp}{%
\begin{equation}\begin{split}
  \BODY
\end{split}\end{equation}
}
\renewcommand{\proofname}{$\because$}


\usepackage{adjustbox}
\usepackage{tabularx}
\newcolumntype{Y}{>{\centering\arraybackslash}X} 
\usepackage{booktabs} 
\usepackage{colortbl} 
\def\beq#1\eeq{\begin{align}#1\end{align}}

\RequirePackage{xspace}
\def\Bbar    {\kern 0.18em\overline{\kern -0.18em B}{}\xspace}

\newcommand{\tk}[1]{\textcolor{orange}{#1}}

\preprint{CHIBA-EP-277, IPMU26-0004}

\title{Full Three-Loop Electroweak Multiplet Contributions to the Electron Electric Dipole Moment}

\author[a]{Tatsuya Banno,}
\author[a,b,c]{Junji Hisano,}
\author[d,b]{Teppei Kitahara,}
\author[a]{Kiyoto Ogawa,}
\author[a]{and\\ Naohiro Osamura}

\affiliation[a]{
  Department of Physics, Nagoya University, Furo-cho Chikusa-ku, Nagoya 464-8602, Japan
}
\affiliation[b]{
  Kobayashi-Maskawa Institute for the Origin of Particles and the
  Universe, Nagoya University,
  Furo-cho Chikusa-ku, Nagoya 464-8602, Japan
}
\affiliation[c]{
  Kavli IPMU (WPI), UTIAS, The University of Tokyo, Kashiwa 277-8584, Japan
}

\affiliation[d]{
Department of Physics, Graduate School of Science,
Chiba University, Chiba 263-8522, Japan}

\emailAdd{banno.tatsuya.p8@s.mail.nagoya-u.ac.jp}
\emailAdd{hisano@eken.phys.nagoya-u.ac.jp}
\emailAdd{kitahara@chiba-u.jp}
\emailAdd{ogawa.kiyoto.f8@s.mail.nagoya-u.ac.jp}
\emailAdd{osamura.naohiro.j2@s.mail.nagoya-u.ac.jp}

\abstract{
Experimental sensitivity to the electric dipole moment (EDM) of the electron has improved remarkably in recent years. 
Consequently, future prospects could probe new physics whose contribution to the electron EDM first arises at three-loop order.
Additional SU(2)$_L$ multiplets with  CP-violating Yukawa interactions, which contribute to the electron EDM at three-loop level, is one such testable new physics scenario.
In this scenario, the electron EDM is radiatively induced from two contributions: the CP-odd trilinear $W$-boson coupling, called the electroweak-Weinberg operator, and the CP-odd dipole operator of electron.
The former and the latter operators are generated at two-loop and three-loop levels, respectively, after integrating out  the SU(2)$_L$ multiplets.
Within the same models, according to an  analysis based on the Standard Model Effective Field Theory (SMEFT), we previously found that the contribution to the electron EDM from the electroweak-Weinberg operator can be probed in future experiments. 
However, the one-loop matching condition between the electron EDM and the electroweak-Weinberg operator does not receive a large logarithmic enhancement because the associated anomalous dimension is zero.
The CP-odd dipole operator of the electron would contribute to the electron EDM  at the same three-loop order as the contribution through the electroweak-Weinberg operator.
In this paper, we directly calculate the electron EDM induced by the CP-violating Yukawa interactions of the SU(2)$_L$ multiplets at \emph{full} three-loop level.
A central result is that the full three-loop calculation is a factor of three larger than that of the electroweak-Weinberg operator alone.
}


\begin{document}
\sloppy 

\makeatletter\renewcommand{\@fpheader}{\ }\makeatother

\maketitle

\renewcommand{\thefootnote}{\#\arabic{footnote}}
\setcounter{footnote}{0}


\section{Introduction}

The electric dipole moment (EDM) of the electron is one of the most sensitive observables for CP violation beyond the Standard Model (BSM).
The ACME~II experiment has achieved a bound of $|d_e| < 1.1 \times 10^{-29} e \text{\ cm}$ using the ThO molecules \cite{ACME:2018yjb}, while the JILA 
experiment has set the current most stringent bound, $|d_e| < 4.1 \times 10^{-30} e\text{\ cm}$, using the HfF$^+$ molecular ions \cite{Roussy:2022cmp}. 
These experimental sensitivities are expected to improve
by several orders of magnitude over the next decades \cite{Alarcon:2022ero}.
As a concrete example, 
the next-generation ACME~III experiment is expected
to implement apparatus upgrades that will improve
the sensitivity by about a factor of $30$ compared to the ACME II, pushing the reach down to $\mathcal{O}(10^{-31}) e\text{\ cm}$ level \cite{Hiramoto:2022fyg}.
Within the Standard Model (SM), the electron EDM induced by the CP-violating phase in the CKM matrix is extremely suppressed, of order $10^{-44}e \text{\ cm}$, since it arises only at the four-loop level \cite{Pospelov:2013sca}. 
The CKM phase also induces the paramagnetic EDM through the semileptonic four-fermion operators, 
for which the predicted effect is equivalent to the electron EDM of  $d^{\rm eq}_e =1.0 \times 10^{-35}e \text{\ cm}$ \cite{Ema:2022yra}.
In contrast, BSM scenarios often contain additional sources of CP violation, leading to the electron EDMs that are much larger than those predicted in the SM. Current experiments are already sensitive to TeV-scale physics that induces the electron EDM at the two-loop level, assuming couplings comparable to the SU(2)$_L$ gauge coupling and the magnitude of the CP-violating phases of $\mathcal{O}(1)$. 
Under these assumptions, future experiments will be able to probe TeV-scale physics that generates the electron EDM even at three-loop level.

Additional SU(2)$_L$ multiplets with CP-violating Yukawa couplings are one of the testable classes of BSMs by using the future electron EDM measurements. 
A primary motivation for studying SU(2)$_L$ multiplet models is dark matter.
If the neutral component of an SU(2)$_L$ multiplet is identified as dark matter and its thermal relic abundance is assumed to reproduce the observed value, the mass of the multiplet is expected to be at the TeV scale \cite{Cirelli:2005uq,Hisano:2006nn,Cirelli:2007xd,Cirelli:2009uv}. 
At the LHC, searches for heavy, non-colored particles such as dark matter are challenging, and the most stringent current lower bound for the SU(2)$_L$  triplet case is 660 GeV~\cite{ATLAS:2022rme}. On the other hand, the electron EDM is sensitive to TeV-scale physics, assuming the magnitude of the CP-violating phases of $\mathcal{O}(1)$. For the SU(2)$_L$ multiplets, the induced EDM can reach the sensitivity of future experiments.

Let us introduce SU(2)$_L$ fermions $\psi_{A/B}$ and a complex scalar $S$ with CP-violating Yukawa couplings among them. 
These Yukawa interactions induce the electron EDM at
three-loop order or higher even when they do not couple directly to the electron at tree
level.\footnote{%
If $S$ is identified with the SM Higgs boson, the electron EDM is generated via the two-loop Barr-Zee diagrams \cite{Barr:1990vd}.
That case has already discussed in Refs.~\cite{Nagata:2014wma,Hisano:2014kua}.}
In Ref.~\cite{Banno:2024apv}, we calculated this contribution based on the effective field theory approach. 
In the Standard Model Effective Field Theory (SMEFT), the CP-odd trilinear $W$-boson coupling, called the electroweak-Weinberg operator, and the CP-odd leptonic dipole operators are generated at two-loop level and three-loop level, respectively, when heavy particles are integrated out.  
We evaluated the electron EDM induced by the electroweak-Weinberg operator through a one-loop matching condition and found that this setup can be probed in future experiments. 
However, this matching contribution is not enhanced by renormalization-group effects because the anomalous dimension vanishes \cite{Jenkins:2017dyc}. 
Therefore, the contribution from the CP-odd electron dipole operator in the SMEFT must be evaluated because both contributions to the electron EDM are the same three-loop order.

In this paper, we evaluate the \emph{full} three-loop contributions to the electron EDM for the case in which $S$
is not the SM Higgs. 
As a result, we find that the electron EDM in the full theory is approximately three times larger than the contribution from the electroweak-Weinberg operator alone.

This paper is organized as follows.
In Sec.~\ref{sec:Yukawa and EFT}, we set up the CP-violating Yukawa interactions of SU(2)$_L$ multiplets and review  Ref.~\cite{Banno:2024apv}, in which the contribution of the electroweak-Weinberg operator to the electron EDM is evaluated. 
In Sec.~\ref{sec:how_to_calculate}, we present the calculation of  the full three-loop contributions to the electron EDM from the CP-violating Yukawa interactions. 
In Sec.~\ref{sec:Numerical_analysis}, we discuss the relationship between the electron EDM in the full theory and the contribution from the electroweak-Weinberg operator, as well as the prospects for the future experiments. 
Section~\ref{sec:conclusion} is devoted to conclusions.

%
\section{Electron EDM via the Electroweak-Weinberg Operator}
\label{sec:Yukawa and EFT}

In this section, we introduce our set up, and review the evaluation of the electron EDM induced by SU(2)$_L$ multiplets with CP-violating Yukawa interactions at three-loop level based on the SMEFT framework.

A CP-violating phase responsible for the electron EDM is introduced through an extended Yukawa interactions,
\begin{eqsp}
    \mathcal{L} 
    \supset
        -
        \bar{\psi}_B g_{\bar{B} A S} \psi_A S
        -
        \bar{\psi}_A g_{\bar{A} B \bar{S}} \psi_B S^*\,,
    \label{eq:Yukawa int}
\end{eqsp}
where the two fermion multiplets ($\psi_A$ and $\psi_B$) and the complex scalar ($S$) are additional heavy particles with any representation of SU(2)$_L$ gauge group.\footnote{For a single fermion multiplet case, 
a CP-violating phase in a Yukawa interaction is removable by rephasing of the complex scalar $S$, and thus does not yield physical CP violation.
On the other hand, for a single-fermion multiplet with a real scalar case,
the CP violation would occur, 
but it entails more diagrams than our study.}
Here, the Yukawa coupling constants are given as
\begin{align}
    g_{\bar{B}AS} &= X_{\bar{B}AS}(s+\gamma_5a)\,, \\
    g_{\bar{A}B\bar{S}} &= X_{\bar{A}B\bar{S}}(s^*-\gamma_5a^*)\,,
\end{align}
where $X_{\bar{B}AS}, X_{\bar{A}B\bar{S}}$ are SU(2)$_L$ invariant tensors which depend on the representation of these fields, and the scalar component $s$ and the pseudoscalar one $a$ are complex numbers in general.
In this set up, the SU(2)$_L$ gauge bosons
can mediate $\psi_{A/B}$ and $S$ to the SM sector.
Hence, the electron EDM is induced by the CP-odd $W W \gamma$ vertex in this model at three-loop level.
We treat the case of $(A,B,S)=(r,r,1)$, where $r$ is for the dimension of SU(2)$_L$ multiplet representation, because it is the simplest case to confirm how large the electron EDM is in the full theory, compared with the contribution of the electroweak-Weinberg operator.
Then, the SU(2)$_L$ invariant tensors are determined as $X_{\bar{B}AS}=X_{\bar{A}B\bar{S}}=\delta^{a_r b_r}$, where $a_r ,b_r$ are indices of the representation for $\psi_{A/B}$.

Observables should be invariant under the basis transformation for the fields in the Lagrangian.
It means that the electron EDM to be composed of CP-odd  rephasing-invariants of parameters in Lagrangian.
In our set up in Eq.~\eqref{eq:Yukawa int}, $\Im(s a^*) m_A m_B$ is the unique CP-odd rephasing-invariant quantity.
Therefore, the contribution to the electron EDM has to be proportional to the factor $\Im(s a^*) m_A m_B$, see the discussion of the rephasing-invariant in Refs.~\cite{Banno:2023yrd,Banno:2025pfq}.

The full three-loop calculation, which is the main subject of this paper, is presented in the next section. In the remainder of this section, we review the evaluation of the electron EDM using the effective field theory approach carried out in Ref.~\cite{Banno:2024apv}.
When we consider the  SMEFT, in  which heavier particles than SM-particles are integrated out in a limit of vanishing the Higgs vacuum expectation value, the CP-odd trilinear SU(2)$_L$ gauge boson operator and the CP-odd leptonic dipole operator are derived as higher dimensional operators related to the electron EDM.
The former called the electroweak-Weinberg operator is generated at two-loop level, and the later is induced at three-loop level.
In Ref.~\cite{Banno:2024apv}, we evaluated the electroweak-Weinberg operator from Eq.~(\ref{eq:Yukawa int}), and derived the matching condition to the electron EDM at one-loop level.

First, we mention the evaluation of the Wilson coefficient for the electroweak-Weinberg operator. 
The electroweak-Weinberg operator is given as
\begin{align}
   \mathcal{L}_W 
    = -\frac{g^3}{3}C_W\epsilon^{abc}W^a_{\mu\nu}W^{b\nu}_{~\ \rho}\widetilde{W}^{c\rho\mu}\,,
    \label{eq:WWW coupling} 
\end{align}
where $g$ is the SU(2)$_L$ gauge coupling, $\epsilon^{abc}$ are the structure constants, and $W^a_{\mu\nu}$ denotes the SU(2)$_L$ field strength.
The dual field strength is defined by $\widetilde{W}^{a\mu\nu} \equiv \frac{1}{2}\varepsilon^{\mu\nu\rho\sigma} W^a_{\rho\sigma}$, with the convention $\varepsilon^{0123}=+1$.
\begin{figure}[t]
    \centering
    \includegraphics[width=0.4\linewidth]{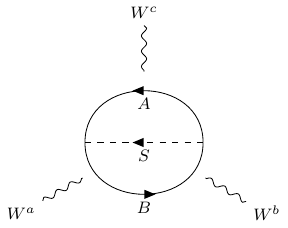}
    \caption{Two-loop diagrams generating the Wilson coefficient $C_W$ of the electroweak-Weinberg operator after integrating out the heavy fields.}
    \label{fig:diagrams_for_Cw}
\end{figure}
The Wilson coefficient $C_W$ can be described as the two-loop diagram in Fig.~\ref{fig:diagrams_for_Cw}.
$C_W$ in the case of $(A,B,S)=(r,r,1)$ is obtained as \cite{Abe:2017sam,Banno:2024apv},
\begin{align}
    C_W &= \frac{6}{(4\pi)^4}  
    \frac{r(r^2 - 1)}{12}
    \Im(sa^*)m_Am_B \nonumber\\
    &\ \times\left[g_1(m_A^2,m_B^2,m_S^2) +   g_2(m_A^2,m_B^2,m_S^2) + (m_A \leftrightarrow m_B)  \right]  \, ,
    \label{eq:Wilson coeffiecient}
\end{align}
where $(m_A \leftrightarrow m_B)$ means the exchange of two fermions mass for $g_1(m_A^2,m_B^2,m_S^2)$ and $   g_2(m_A^2,m_B^2,m_S^2)$.
The loop functions $g_1$ and $g_2$ are given by
\begin{align}
    g_1(x_1,x_2,x_3) &= \left( 2\bar{I}_{(4;1)} + 4x_1\bar{I}_{(5;1)} \right)(x_1;x_2;x_3)\,, \\
    g_2(x_1,x_2,x_3) &= \left( \bar{I}_{(3;2)} + x_1\bar{I}_{(4;2)} \right)(x_1;x_2;x_3)\,,
\end{align}
with $\bar{I}_{(n;m)}$ defined in terms of the UV-finite master integral $\bar{I}(x_1;x_2;x_3)$ as
\begin{align}
    \bar{I}_{(n;m)}(x_1;x_2;x_3)
    = \frac{1}{(n-1)!(m-1)!} \frac{d^{n-1}}{dx_1^{n-1}} \frac{d^{m-1}}{dx_2^{m-1}} \bar{I}(x_1;x_2;x_3)\,.
\end{align}
Analytic expressions for $\bar{I}(x_1;x_2;x_3)$ are found in Refs.~\cite{Ford:1992pn,Espinosa:2000df,Martin:2001vx}.
Since the CP-violating phase enters through the relative phase between the scalar and pseudoscalar Yukawa couplings, the result is proportional to $\mathrm{Im}(s a^*)$.
Moreover, chirality requires a mass insertion on each fermion line, yielding an overall factor $m_A m_B$.
Therefore, product between $\mathrm{Im}(s a^*)$ and $m_A m_B$ composes of the rephasing-invariance for Eq.~\eqref{eq:Wilson coeffiecient}.

\begin{figure}
    \begin{center}
        \includegraphics[width = 0.45 \textwidth]{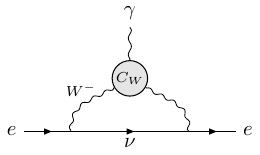}
        \caption{
             One-loop matching diagram that induces the electron EDM from the electroweak-Weinberg operator. 
 }
            \label{fig:EDM through CW}
    \end{center}
\end{figure}

The electron EDM can be obtained from the electroweak Weinberg operator through a one-loop matching diagram, as shown in  Fig.~\ref{fig:EDM through CW}.
This calculation exhibits regularization dependence \cite{Jenkins:2017dyc,Jenkins:2017jig}. In dimensional regularization using the BMHV scheme \cite{tHooft:1972tcz,Breitenlohner:1977hr}, both the four-dimensional part and the evanescent operator associated with the electroweak Weinberg operator contribute to the matching condition. Here, the evanescent operator denotes an operator that vanishes in the four-dimensional limit. The matching condition for the four-dimensional part in dimensional regularization was evaluated in Refs.~\cite{Hoogeveen:1987jn,Atwood:1990cm,Boudjema:1990dv,DeRujula:1990db,Novales-Sanchez:2007rsw}. We then showed that the result of Ref.~\cite{Boudjema:1990dv} is correct only for the four-dimensional part within dimensional regularization.

The matching condition for four-dimensional part is obtained as,
\begin{equation}
    \frac{d_e^{C_W}}{e}
    =
        \frac{1}{6}
        \left(\alpha_2\right)^2 m_e 
        C_W 
        +
        \mathcal{O} \left(\frac{m_e^2}{m_W^2}\right)\,,
    \label{eq:matching}
\end{equation}
where $\alpha_2 = g^2/4\pi$ and $m_W$ is the $W$ boson mass.
In terms of the evanescent contribution, we confirmed that the matching condition is consistent with Ref.~\cite{Dekens:2019ept,Abe:2024mwa} when the Wilson coefficients for the four-dimensional and evanescent parts are common.
However, it is nontrivial that the evanescent operator for the electroweak-Weinberg operator is equal to $C_W $.
Then, we treat only the four-dimensional part in Eq.~\eqref{eq:matching} as the contribution of the electron EDM induced by the electroweak-Weinberg operator.
Contrary to expectation, this result does not diverge, and hence the logarithmic term does not appear.
Therefore, the operator mixing to the electron EDM from the electroweak-Weinberg operator does not occurred through renormalization-group equation \cite{Jenkins:2017dyc}, and $d_e^{C_W}$ is composed from only the threshold correction in the low-energy effective field theory (LEFT) below the electroweak scale.

We can finally obtain the induced electron EDM by combining Eqs.~\eqref{eq:Wilson coeffiecient} and ~\eqref{eq:matching}. 
On the other hand,  
the CP-violating leptonic dipole operator in the SMEFT also directly contributes to the electron EDM. Since the matching contribution to the electron EDM from the electroweak Weinberg operator does not receive any logarithmic enhancement, the CP-violating leptonic dipole operator contributes at the same three-loop order as Eq.~\eqref{eq:matching}. Thus, a full three-loop calculation of the electron EDM is required in order to obtain an accurate prediction.
\section{Full Three-Loop Calculations}
\label{sec:how_to_calculate}
\begin{figure}[htb]
  \centering
  \begin{tabular}{ccc}
    \includegraphics[width=0.25\linewidth]{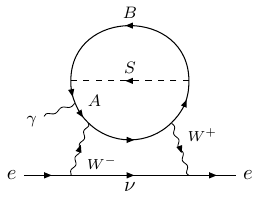} &
    \includegraphics[width=0.25\linewidth]{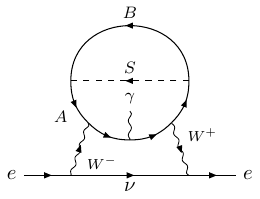} &
    \includegraphics[width=0.25\linewidth]{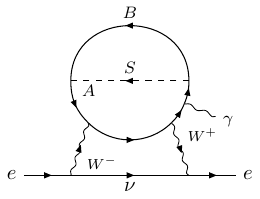} \\
    
    \includegraphics[width=0.25\linewidth]{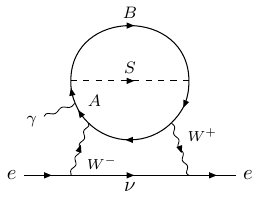} &
    \includegraphics[width=0.25\linewidth]{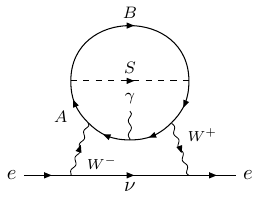} &
    \includegraphics[width=0.25\linewidth]{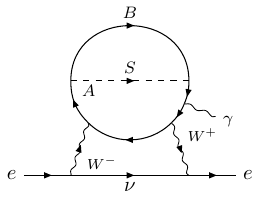} \\

    \includegraphics[width=0.25\linewidth]{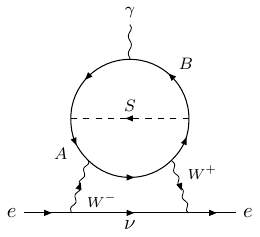} &
    \includegraphics[width=0.25\linewidth]{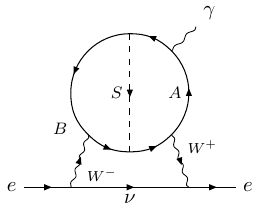} &
    \includegraphics[width=0.25\linewidth]{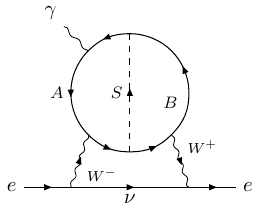} \\
    
    \includegraphics[width=0.25\linewidth]{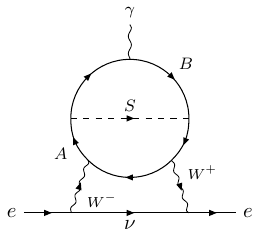} &
    \includegraphics[width=0.25\linewidth]{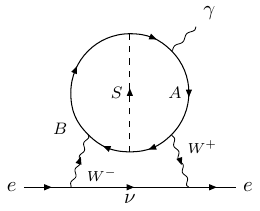} &
    \includegraphics[width=0.25\linewidth]{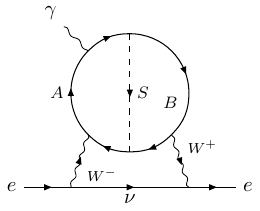}
  \end{tabular}
  \caption{Feynman diagrams contribute to the electron EDM at three-loop level.
  }
  \label{fig:Feynman_Diagrams}
\end{figure}
In this section, we explain how to calculate full three-loop diagram in the case of the representation of SU(2)$_L$ multiplets $(A,B,S)=(r,r,1)$.
Because the complex scalar $S$ is SU(2)$_L$ singlet, there are twelve Feynman diagrams contributing to the electron EDM in the full theory as Fig.~\ref{fig:Feynman_Diagrams}.
Each diagram has corresponding diagrams obtained by exchanging $\psi_A$ and $\psi_B$, as well as the Yukawa interactions.

In this set up, the three-loop diagrams involving the $WW$ and $ZZ$ two-point functions at two-loop level could, in principle,  generate the electron EDM. 
However, we confirm that such diagrams do not contribute the electron EDM because the two-point functions cannot generate the Levi-Civita symbol in the calculation of the electron EDM. It is also found that the $ZZ\gamma$ and $\gamma\gamma\gamma$ vertices at the two-loop level do not contribute to the electron EDM since the clockwise and anticlockwise fermion loops are canceled out by each other. Therefore, only the Feynman diagrams shown in Fig.~\ref{fig:Feynman_Diagrams} contribute to the electron EDM.



The contribution to the electron EDM in the full theory is given by 
\begin{align}
    \frac{d^{\rm Full}_e}{e}
    =
    \frac{\alpha_2^2}{(16 \pi^2)^2}
    \frac{m_e}{2}
    \frac{r (r^2-1)}{12}
    \Im (s a^*)
    m_A m_B  
    B(m_A,m_B,m_S,m_W)\,,
    \label{eq:eEDM_Full}
\end{align}
where $B(m_A,m_B,m_S,m_W)$ denotes a combination of three-loop integrals.
As a consistency check of Eq.~(\ref{eq:eEDM_Full}), when two $W$ bosons satisfy the on-shell condition, we confirm that the part of the effective vertex $WW\gamma$ at two-loop level is consistent with the Wilson coefficient of the electroweak-Weinberg operator in Eq.~(\ref{eq:Wilson coeffiecient}). 
Since we consider the case of the heavier SU(2)$_L$ multiplets than the $W$ boson mass, $m_W$, $B(m_A,m_B,m_S,m_W)$ can be expanded as 
\begin{align}
    B(m_A,m_B,m_S,m_W)
    \simeq
    B_0(m_A,m_B,m_S)
    +
    m^2_W B_1(m_A,m_B,m_S)
    +
    \mathcal{O} \left( \frac{m^4_W}{\Lambda^4} \right)\,,
    \label{eq:B}
\end{align}
where $\Lambda$ stands for the heavy particle scale ($\psi_{A/B}$ or $S$).
$B_0(m_A,m_B,m_S)$ and $ B_1(m_A,m_B,m_S)$ are sum of the three-loop integrals defined as
\begin{align}
    &J[n_1,n_2,n_3,n_4,n_5,n_6] 
    =
    \left(\frac{i}{16 \pi^2}\right)^{-3}
    \int \frac{d^d l_1}{(2 \pi)^{d}} \frac{d^d l_2}{(2 \pi)^{d}} \frac{d^d l_3}{(2 \pi)^{d}}
    \nonumber
    \\
    &
    \frac{1}
    {(l^2_1-m_A^2)^{n_1} 
    (l^2_2-m_B^2)^{n_2}
    (l^2_3)^{n_3}
    [(l_1-l_2)^2 - m^2_S]^{n_4}
    [(l_2-l_3)^2 - m_B^2]^{n_5}
    [(l_3-l_1)^2 - m_A^2]^{n_6}
    }\,.
    \label{eq:vacuum_int}
\end{align}
The results of $B_0(m_A,m_B,m_S)$ and $B_1(m_A,m_B,m_S)$ are shown in Appendix \ref{app:res_B0_B1}.
Here, $J[n_1,n_2,n_3,n_4,n_5,n_6]$ must be defined within the dimensional regularization ($d=4-2 \epsilon$), because some of the $J[n_1,n_2,n_3,n_4,n_5,n_6]$ that constitute  $B_0(m_A,m_B,m_S)$ and $B_1(m_A,m_B,m_S)$ are  divergent.
However, we confirmed that $B_0(m_A,m_B,m_S)$ and $B_1(m_A,m_B,m_S)$ are finite due to cancellation of the UV and IR divergences.

We make two comments about Eq.~(\ref{eq:eEDM_Full}).
First,  this contribution is rephasing invariant for the same reasons as Eq.~\eqref{eq:Wilson coeffiecient}.
The part including a $\gamma_5$ in the fermion trace is proportional to $\Re(s a^*)$ or $\Im(s a^*)$.
However, the part of the $\Re(s a^*)$ does not contribute to the electron EDM because the corresponding loop-integral vanishes.
As a result, the electron EDM is proportional to only the part of the $\Im(s a^*)$.
Also, fermion masses $m_A$ and $m_B$ are picked up through  chirality flips.
Therefore, the electron EDM is proportional to $\Im(s a^*) m_A m_B$, which is manifestly rephasing invariant.

Second, Eq.~(\ref{eq:eEDM_Full}) is proportional to a factor of $r (r^2 - 1)/12$.
The sum of all diagrams contributing to the $\Im(sa^*)$ part is proportional to $\Tr[Q[T^+ , T^-]] $\,.
Here, $T^\pm,Q$ are composed of SU(2)$_L$ generators and U(1)$_Y$ shown as $T^\pm = T^1 \pm i T^2,\; Q=T^3 + Y$.
By using the group identity, we obtain the equation given as 
\begin{align}
& \Tr\left[Q[T^+ , T^-]\right]=\frac{r (r^2 - 1)}{12}\,.
\end{align}
This result is independent of  U(1)$_Y$  because of $\Tr[T^a] =0$.

To evaluate $B_0(m_A,m_B,m_S)$ and $B_1(m_A,m_B,m_S)$, we employ the integration-by-parts (IBP) method. The
IBP method reduces the complex loop-integrals to linear combinations of the master-integrals.
We perform the IBP reduction of $B_0(m_A,m_B,m_S)$ and $B_1(m_A,m_B,m_S)$ using the public code \texttt{Kira} \cite{Maierhofer:2017gsa, Klappert:2020nbg, Lange:2025fba} together with \texttt{Fermat} \cite{Fermat}.
After the IBP method, we evaluate $B_0(m_A,m_B,m_S)$ and $ B_1(m_A,m_B,m_S)$ using the corresponding master integrals \cite{Martin:2016bgz}.

In the case of $m_A = m_B$, $B_0(m_A,m_B,m_S)$ and $B_1(m_A,m_B,m_S)$ can be expressed in terms of analytic functions.
In this degenerate-mass case,  $B_0(m_A,m_B,m_S)$ and $ B_1(m_A,m_B,m_S)$ can be reduced to seven master-integrals as follows: 
\begin{align}
\{
     &J[1,0,0,1,1,0],~J[1,1,0,0,1,0],~J[1,1,0,1,1,0],
     \nonumber
     \\
     &J[0,1,1,2,0,1],~J[0,2,1,1,0,1],~J[2,1,0,0,1,1],~J[1,1,0,1,1,1] 
    \}\,.
    \label{eq:three-loop_list}
\end{align}
The integrals $J[1,0,0,1,1,0]$ and $J[1,1,0,0,1,0]$ are combinations of one-loop vacuum integrals, while $J[1,1,0,1,1,0]$ is given by the product of one-loop vacuum integral and two-loop one. 
The integrals $J[0,1,1,2,0,1],~J[0,2,1,1,0,1],~J[2,1,0,0,1,1]$, and $J[1,1,0,1,1,1]$ are genuine three-loop vacuum integrals.
These master integrals are given analytically in Refs.~\cite{Martin:2016bgz,Martin:2001vx}.
However, the resulting expressions for $B_0(m_A,m_B,m_S)$ and $B_1(m_A,m_B,m_S)$ still remain rather complicated.
Therefore, we provide the analytic result for the contribution to the electron EDM  in the ancillary file \texttt{solB.txt}.\footnote{In the case of  $m_A \neq m_B \neq m_S$, $B_0(m_A,m_B,m_S)$ and $B_1(m_A,m_B,m_S)$ can only be evaluated numerically.
In this case, it is convenience to use \texttt{3VIL}~\cite{Martin:2016bgz} which is a public-code to calculate three-loop vacuum integrals after the IBP method.}

\section{Numerical Analysis}
\label{sec:Numerical_analysis}

In this section, we evaluate the electron EDM induced by the CP-violating Yukawa interactions in Eq.~(\ref{eq:Yukawa int}) at the full three-loop level, and compare it with the contribution mediated by the electroweak-Weinberg operator in Eq.~\eqref{eq:matching}. 
We also discuss the prospects for the future experiments.
To obtain analytic results, we consider the case of degenerate fermion masses. 
First, we study the simplest case of  $m_A=m_B=m_S$.
Second, we discuss the case of $m_A=m_B\neq m_S$.

\subsection{\texorpdfstring{$m_A=m_B=m_S$}{mA = mB = mS}}

In the case of $m_A = m_B = m_S$, the electron EDM in Eq.~(\ref{eq:eEDM_Full}) induced by the CP-violating Yukawa interactions in Eq.~(\ref{eq:Yukawa int}) at full three-loop level is obtained as follows:
\begin{align}
    \frac{d^{\rm Full}_e}{e}
    &=
    \frac{\alpha_2^2 m_e}{(16 \pi^2)^2}
    \frac{r (r^2-1)}{12}
    \Im (s a^*)
    \nonumber
    \\
    & \quad \times
    \left[
    \frac{-1 + 4 \sqrt{3} {\rm Cl_2} (2\pi/3)}{9 m_A^2}
    +
    \frac{(-209 + 224 \sqrt{3} {\rm Cl_2} (2\pi/3))m^2_W}{243 m_A^4}
    \right]
    +
    \mathcal{O}\left(\dfrac{m_W^4}{m_A^4}\right)
    \nonumber
    \\
    &=
    \frac{\alpha_2^2 m_e}{(16 \pi^2)^2}
    \frac{r (r^2-1)}{12}
    \Im (s a^*)
    \left(
    \frac{0.41}{m_A^2}
    +
    \frac{0.22m^2_W}{m_A^4}
    \right)
    +
   \mathcal{O}\left(\dfrac{m_W^4}{m_A^4}\right)\,,
    \label{eq:EDM_Full_same_mass}
\end{align}
where ${\rm Cl_2}(x)$ is known as the Clausen function of order 2.
On the other hand, the contribution through the electroweak-Weinberg operator in Eq.~(\ref{eq:matching}) is given as \cite{Banno:2024apv}
\begin{align}
    \frac{d^{C_W}_e}{e}
    &=
    \frac{\alpha_2^2 m_e}{(16 \pi^2)^2}
    \frac{r (r^2-1)}{12}
    \Im (s a^*) 
    \times
    \frac{-1 + 4 \sqrt{3}\rm{Cl}_2 (2\pi/3)}{27 m_A^2}
    \nonumber
    \\
    &=
    \frac{\alpha_2^2 m_e}{(16 \pi^2)^2}
    \frac{r (r^2-1)}{12}
    \Im (s a^*) 
    \frac{0.14}{m_A^2}\,.
    \label{eq:EDM_EFT_same_mass}
\end{align}
When we specifically compare Eq.~\eqref{eq:EDM_Full_same_mass} excluding the next-to-leading-order (NLO) terms,  with Eq.~(\ref{eq:EDM_EFT_same_mass}), we found that the full contribution is exactly three times larger, independent of the representation.

\begin{figure}[t]
    \centering
    \includegraphics[width=0.5\linewidth]{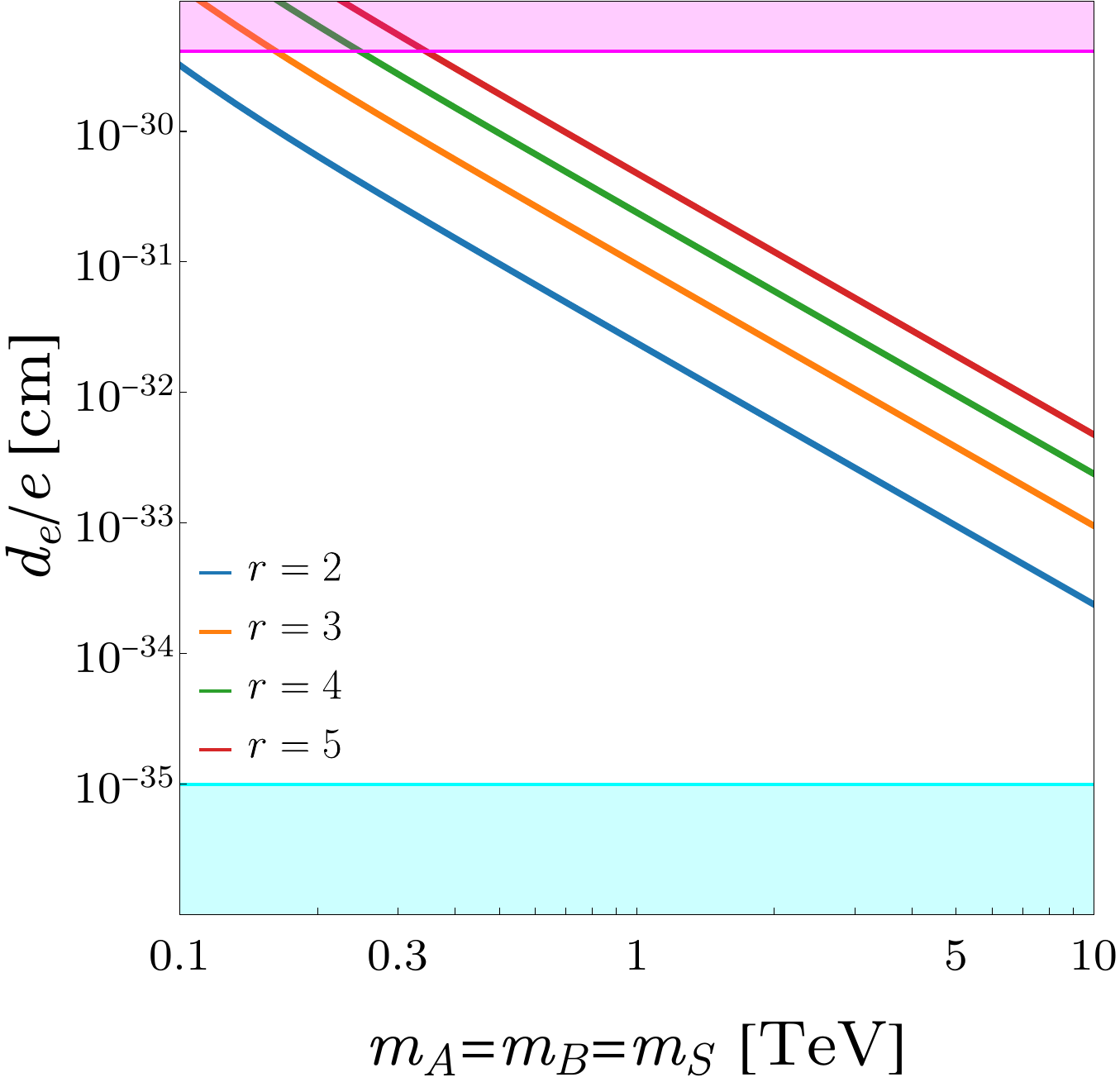}
    \caption{Electron EDM induced by the CP-violating Yukawa couplings of $(r,r,1)$ SU(2)$_L$ multiplets degenerated mass $m_A=m_B=m_S$ in full theory. Here,  $\opn{Im} (s a^*) = 0.25$ is taken.
    The four color lines illustrate each SU(2)$_L$ representation, where $r=2,3,4,5$, respectively. 
    The magenta region is excluded by current experimental bound, $|d_e^{\rm exp.}| < 4.1 \times 10^{-30} e \, {\rm cm}$ \cite{Roussy:2022cmp}, and the cyan region cannot be probed due to CKM contribution through $e$--$N$ four-Fermi interaction~\cite{Ema:2022yra}.
    }
    \label{fig:b_1flavor}
\end{figure}

Figure~{\ref{fig:b_1flavor}} shows whether this scenario can be probed in future experiments.
The four color lines correspond to different  SU(2)$_L$ representation with $r=2,3,4$ and $5$, respectively.
We fix the Yukawa coupling at $\opn{Im} (s a^*) = 0.25$ and use this value throughout the following figures as a reference value. The other physical inputs are the electron mass and the SU(2)$_L$ coupling constant $\alpha_2 = 0.034$ \cite{PDG:2024}.
The magenta region is excluded by the current experimental bound on the electron EDM ($|d^{\rm exp}_e| < 4.1 \times 10^{-30}e$\,cm) \cite{Roussy:2022cmp}. 
The cyan-shaded region indicates that the electron EDM induced by the Yukawa interactions is  smaller than the the SM contribution, where 
 the dominant contribution to paramagnetic atomic or molecular EDMs arises from the CKM phase through semileptonic four-Fermi operators, yielding $d^{\rm eq}_e=1.0\times 10^{-35}e$\,cm \cite{Ema:2022yra}.

As shown in Fig.~{\ref{fig:b_1flavor}}, 
$d^{\rm Full}_e$ is significantly enhanced 
as the SU(2)$_L$ representation dimension $r$ increases,
since Eq.~(\ref{eq:eEDM_Full}) is proportional to $r (r^2-1)$.  
In the minimal dark matter models, larger SU(2)$_L$ representations are often favored to ensure the stability of dark matter. A well-known example is the fermionic quintuplet ($r=5$), which serves as a viable fermionic dark matter candidate \cite{Cirelli:2005uq}. 
In this case, the mass of the quintuplet fermion is typically favored to lie below about 10\,TeV in order to reproduce the observed thermal relic abundance of dark matter \cite{Cirelli:2007xd,Cirelli:2009uv} .
For $r=5$, the mass region with $m_A = m_B =m_S \lesssim 350$~GeV is already excluded by current electron EDM experiment.
Future experiments are expected to achieve sensitivity improvements of at least an order of magnitude.
Therefore, the electron EDM induced by the Yukawa interactions might be probed until the TeV-scale mass region.

\subsection{\texorpdfstring{$m_A = m_B \neq m_S$}{mA = mB neq mS}}
In this section, we discuss the case of $m_A = m_B \neq m_S$.
Even if the scalar mass does not degenerate with the fermion masses, Eq.~(\ref{eq:eEDM_Full}) can be evaluated analytically.
However, since it is difficult to write on this paper, we give Eq.~(\ref{eq:B}) as \texttt{solB.txt}.
When the heavy particles exhibit a large mass hierarchy, the electron EDM in the full theory $d^{\rm Full}_e$ can be expressed as
\begin{align}
    \frac{d^{\rm Full}_e}{e}
    \simeq
     \left\{
        \begin{aligned}
                -&
    \frac{\alpha_2^2 m_e}{(16 \pi^2)^2}
    \frac{r (r^2-1)}{12}
    \Im (s a^*)\times
    \left(
    \frac{3 + 2 \log{\frac{m_A^2}{m^2_S}}}{m^2_S}
    +
    m^2_W
    \frac{7+6 \log{\frac{m_A^2}{m^2_S}}}{9 m_A^2 m^2_S}
    \right) 
    & ( m_A = m_B \ll m_S ) \,,
                                \\
    &
    \frac{\alpha_2^2 m_e}{(16 \pi^2)^2}
    \frac{r (r^2-1)}{12}
    \Im (s a^*)\times
    \left(
    \frac{1}{2m_A^2}
    +
    m^2_W
    \frac{8}{27 m_A^4}
    \right)
    & ( m_S \ll m_A = m_B  ) \,.
        \end{aligned}
     \right.
     \label{eq:EDM_Full_Series}
\end{align}
On the other hands, the electron EDM through the electroweak-Weinberg operator is given as \cite{Banno:2024apv}
\begin{align}
    \frac{d^{C_W}_e}{e}
    \simeq
     \left\{
        \begin{aligned}
                -&
    \frac{\alpha_2^2 m_e}{(16 \pi^2)^2}
    \frac{r (r^2-1)}{12}
    \Im (s a^*)\times
    \left(
    \frac{3 + 2 \log{\frac{m_A^2}{m^2_S}}}{3 m^2_S}
    \right) 
    & ( m_A = m_B \ll m_S ) \,,
                                \\
    &
    \frac{\alpha_2^2 m_e}{(16 \pi^2)^2}
    \frac{r (r^2-1)}{12}
    \Im (s a^*)\times
    \left(
    \frac{1}{6 m_A^2}
    \right)
    & (  m_S \ll m_A = m_B ) \,.
        \end{aligned}
     \right.
    \label{eq:EDM_EFT_Series}
\end{align}
Again, we find that the full contribution $d^{\rm Full}_e$ is exactly three times larger than the electroweak-Weinberg contribution $d^{C_W}_e$, if one ignores the NLO terms.

The logarithmic contributions in ${d^{\rm Full}_e}/{e}$ and ${d^{C_W}_e}/{e}$ in the limit of large $m_S$ originate from 
the renormalization-group effect of EDMs for $\psi_{A/B}$, which is generated by the four-Fermi operators of $\psi_{A/B}$. After integrating out  $\psi_{A/B}$, the EDMs generates contributions to the electroweak-Weinberg operator at one-loop level.
The proportionality of ${d^{\rm Full}_e}/{e}$ and ${d^{C_W}_e}/{e}$ could be understood by evaluating the threshold correction, that does not seem to contain any other logarithmic enhancement terms, for the electron EDM from the EDMs of $\psi_{A/B}$ at two-loop level.

\begin{figure}[t]
    \centering
    \includegraphics[width=0.5\linewidth]{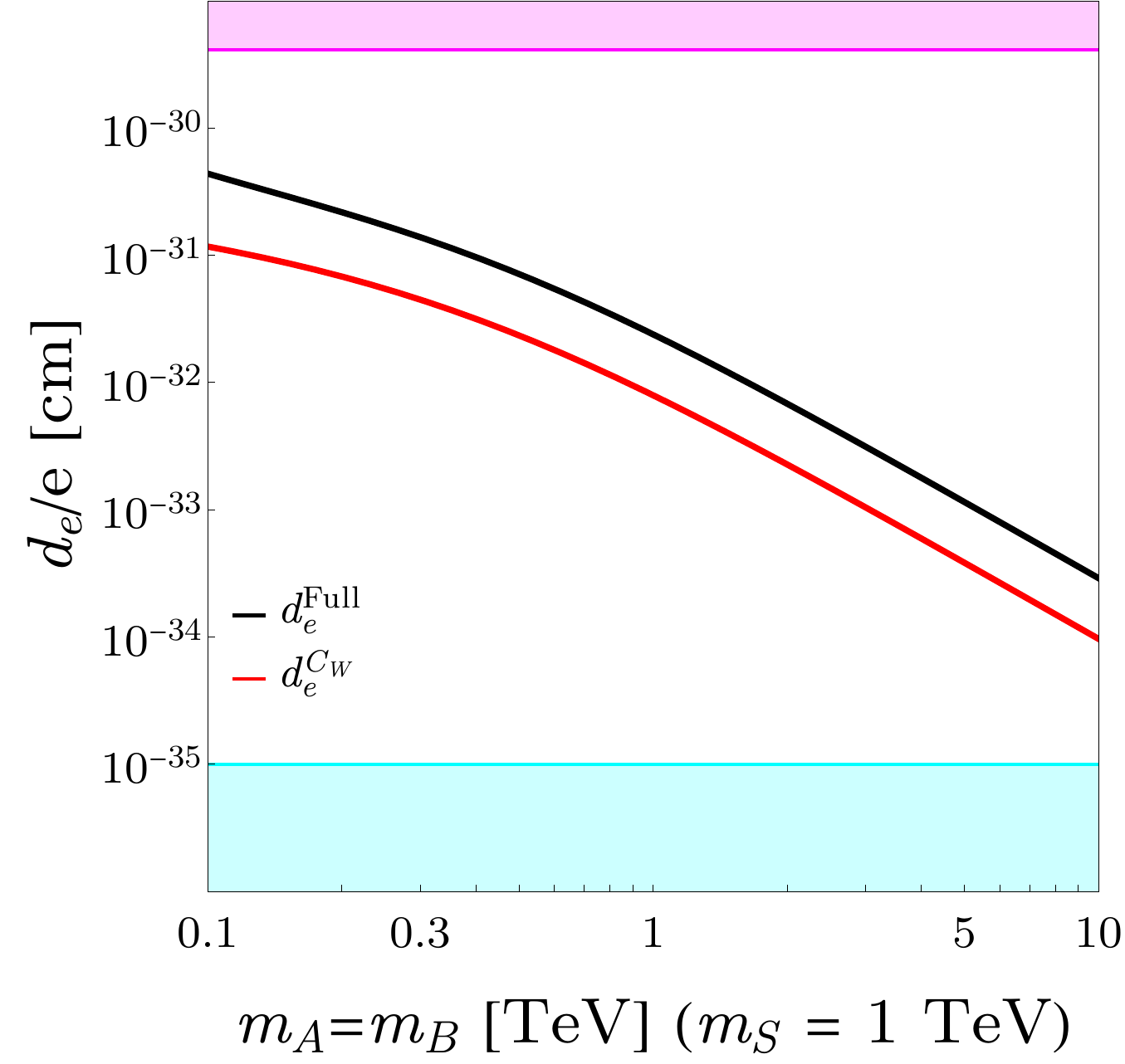}
    \caption{
    Comparison of contributions for the electron EDM in the full theory and only through the electroweak-Weinberg operator when  representations of two fermions are $r=2$.
    Here, $m_S$ is fixed to 1~TeV, and $\opn{Im} (s a^*) = 0.25$ is taken.
    Black (Red) solid line shows the contribution to the electron EDM in the full theory (effective field theory for the electroweak-Weinberg operator).
    }
    \label{fig:2flavor_Full_EFT}
\end{figure}

Figure~\ref{fig:2flavor_Full_EFT} shows a comparison between the electron EDM contributions in the full theory and those mediated by the electroweak Weinberg operator for the case in which the two fermions belong to the  $r=2$ representation.
Here, the scalar mass is fixed to  $m_S=$1~TeV, and the other parameters are taken to be the same as in Fig.~\ref{fig:b_1flavor}.
We find that $d^{\rm Full}_e$ is larger than $d^{C_W}_e$.

\begin{figure}[t]
    \centering
    \includegraphics[width=0.5\linewidth]{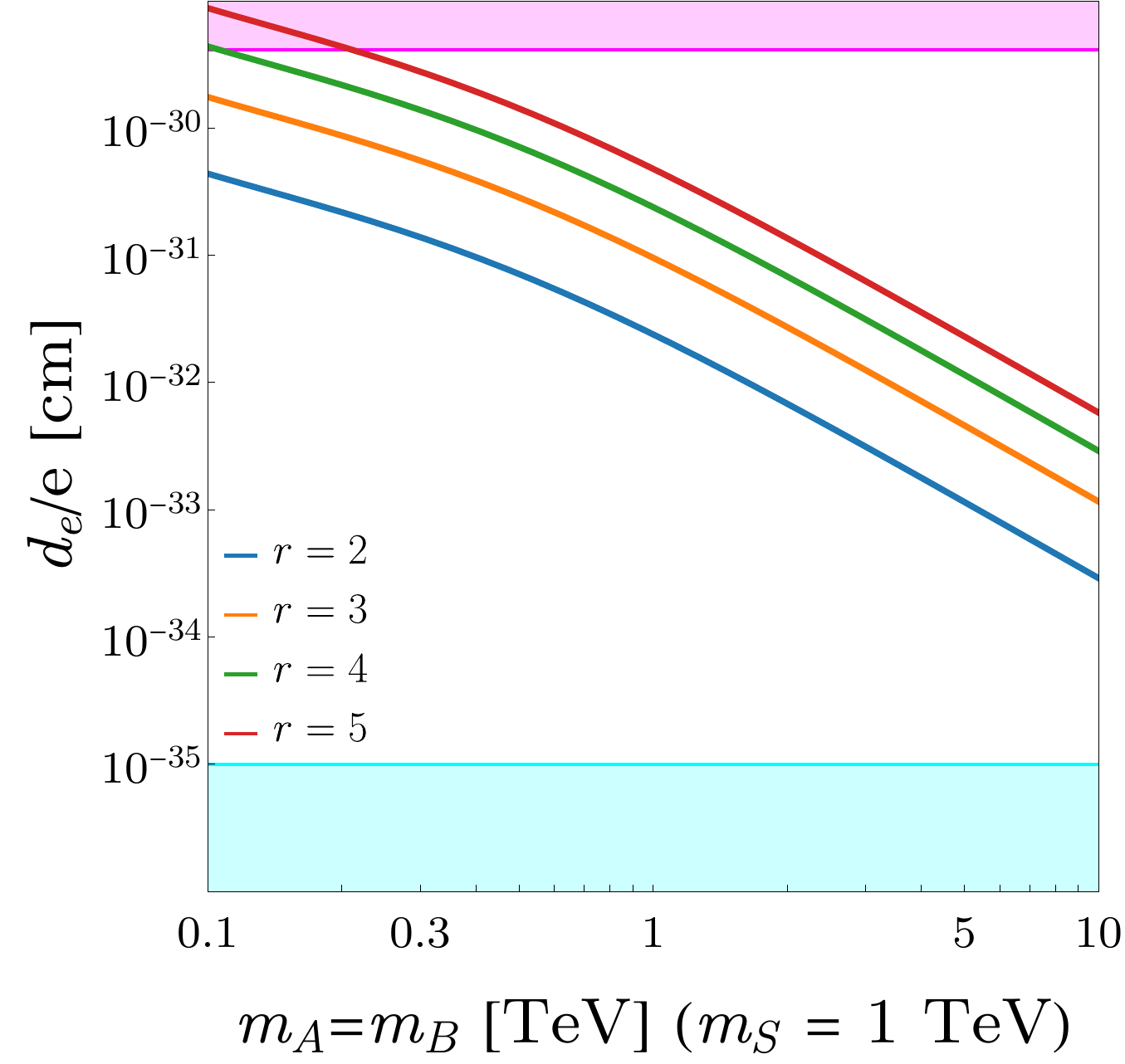}
    \caption{Electron EDM induced by the Yukawa couplings of $(r,r,1)$ SU(2)$_L$ multiplets with $m_A=m_B\neq m_S$ in the full theory. 
    Here, $m_S$ is fixed to  1~TeV, and $\opn{Im} (s a^*) = 0.25$ is taken.
    The four colored lines illustrate each SU(2)$_L$ representation, where $r=2,3,4,5$, respectively. }
    \label{fig:2flavor_1TeV}
\end{figure}

\begin{figure}[t]
    \centering
            \begin{tabular}{cc}
                \begin{subfigure}{0.45 \textwidth}
                    \includegraphics[width = \textwidth]{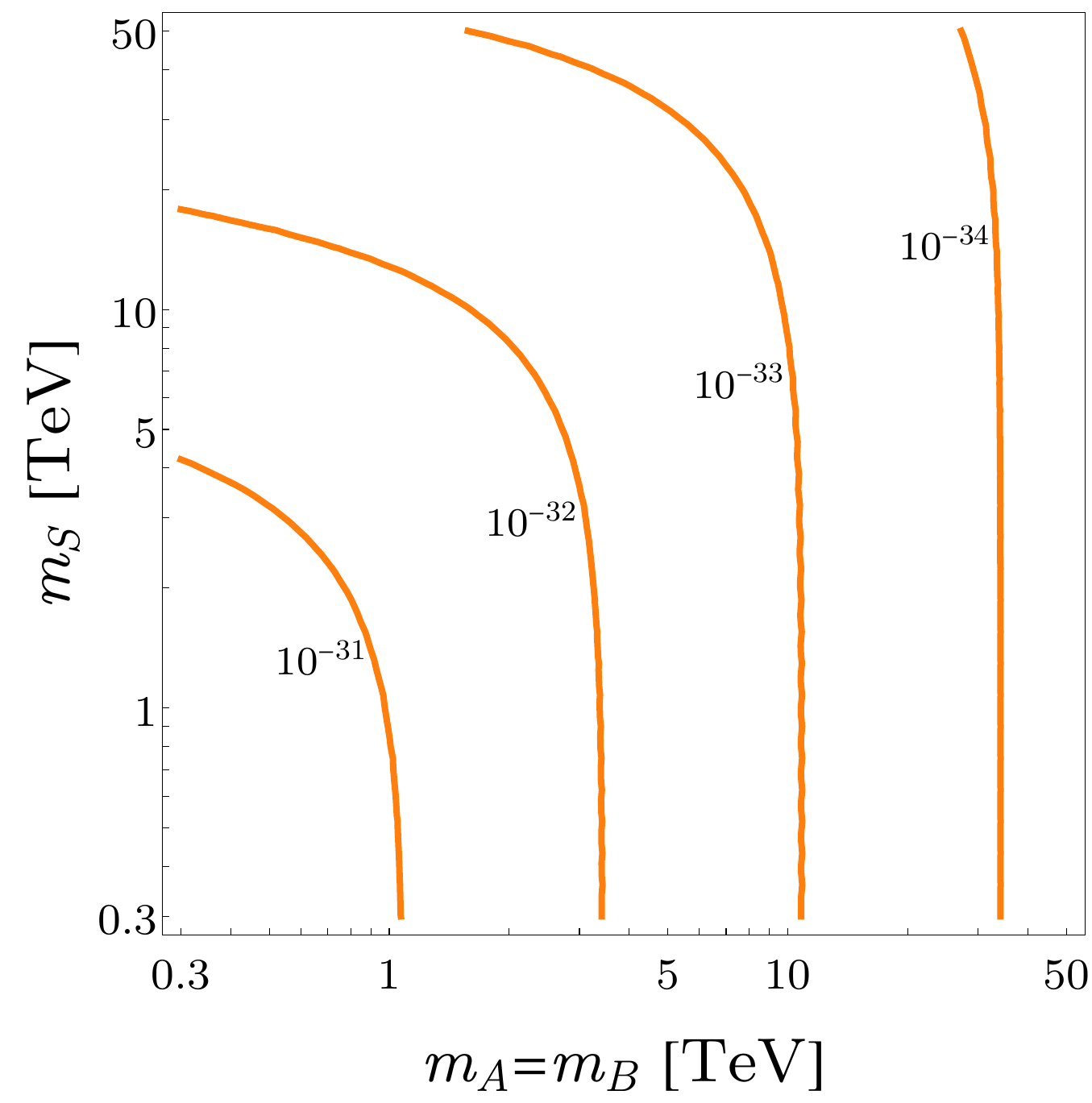}
                    \caption{
                    $r=3$    
                    }
                    \label{fig:2flavor_3}
                \end{subfigure}
                &
                \begin{subfigure}{0.45 \textwidth}
                    \includegraphics[width = \textwidth]{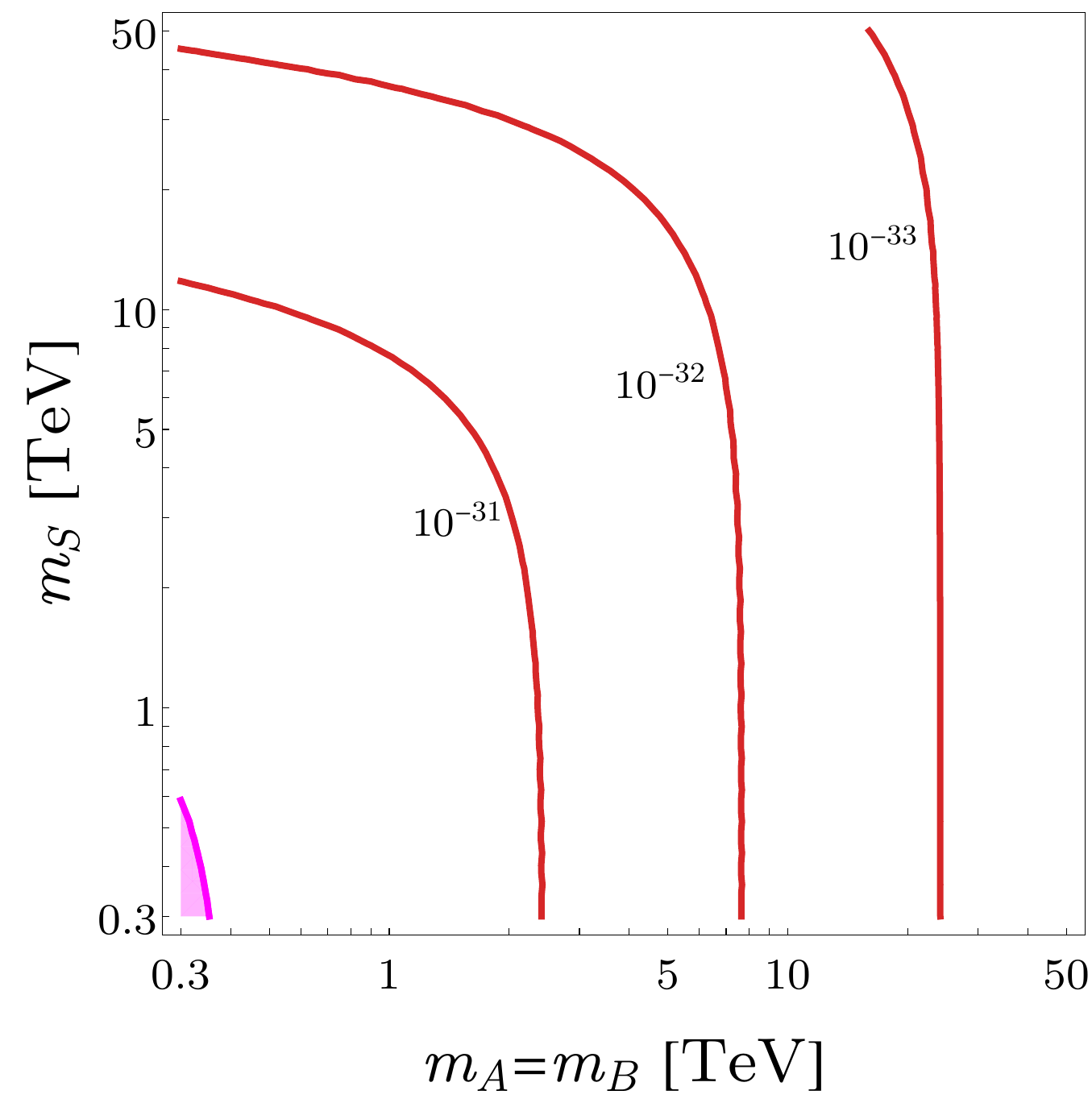}
                    \caption{
                    $r=5$ 
                    }
                    \label{fig:2flavor_5}
                \end{subfigure}
            \end{tabular}
    \caption{
    Contour plots for $d_e^{\rm Full}$ in the cases of $r=3$ and $r=5$. 
    Here, $\opn{Im} (s a^*) = 0.25$ is taken.}
    \label{fig:2flavor_contour}
\end{figure}

Figures~\ref{fig:2flavor_1TeV} and  \ref{fig:2flavor_contour} show the result of the electron EDM induced by Yukawa interaction at full theory in the case of $m_A=m_B \neq m_S$.
In Fig.~\ref{fig:2flavor_1TeV}, the scalar mass is fixed to 1~TeV.
The four color lines illustrate each SU(2)$_L$ representation, where $r=2,3,4,5$, respectively. 
Figures~\ref{fig:2flavor_3} and \ref{fig:2flavor_5} are contour plots in the cases of $r=3$ and $r=5$, with varying the value of $m_S$.
As shown in Fig.~\ref{fig:2flavor_1TeV} and  Fig.~\ref{fig:2flavor_contour}, if the upper limit of the electron EDM can be improved, TeV-scale is expected to cover.

\section{Conclusions}
\label{sec:conclusion}

The SU(2)$_L$ multiplets with  CP-violating Yukawa interactions at three-loop level constitute a testable class of BSM scenarios in the future electron EDM experiments.
In Ref.~\cite{Banno:2024apv}, we previously evaluated the contribution of the electron EDM through the electroweak-Weinberg operator, and found that this set up can be probed.
However, the one-loop matching condition to the electron EDM from the electroweak-Weinberg operator has no renormalization-group effect because the anomalous dimension is zero.
Then, the threshold correction, namely, the contribution from the CP-violating leptonic dipole operator in the SMEFT, contributes at the same three-loop order as Eq.~\eqref{eq:matching}.

In this paper, we evaluated the electron EDM induced by the CP-violating Yukawa interactions of SU(2)$_L$  multiplets at the full three-loop level. As a result, we find that the full calculation yields an electron EDM that is approximately three times larger than the contribution mediated by the electroweak- Weinberg operator. In particular, the fermionic quintuplet case ($r=5$), motivated by the Minimal Dark Matter models, induces a sizable electron EDM because the contribution is enhanced by the cubic power of the SU(2)$_L$ representation dimension. Therefore, this scenario can be probed in future experiments.

In this work, we focused on the specific case $(A,B,S)=(r,r,1)$, which provided a simple setup to assess the magnitude of the electron EDM contribution in the full theory. We did not consider the most general case of SU(2)$_L$ multiplets with arbitrary representations. In such a general scenario, one would need to evaluate additional Feynman diagrams involving gauge-boson insertions on the scalar lines of the SU(2)$_L$ multiplets. We leave this more general analysis for future work.

\acknowledgments
We are sincerely grateful to Yukinari Sumino
for providing us with intensive lectures on the details of higher-order loop calculations.
We would also like to thank  Ayami Hiramoto and Tamaki Yoshioka for useful discussions.
This work is supported by the JSPS Grant-in-Aid for Scientific Research Grant No.\,24K07016 (J.H.), No.\,25H02180 (J.H.), No.\,24K22872 (T.K.), and No.\,25K07276 (T.K.). 
The work of J.H.\ is also supported by 
World Premier International Research Center Initiative (WPI Initiative), MEXT, Japan.
The work of N.O. is supported by JSPS KAKENHI Grant Number 24KJ1256.
T.B. and K.O. are supported by the ``Make New Standards Program for the Next Generation Researchers'' of the Tokai National Higher Education and Research System (THERS).
This work was also financially supported by JST SPRING, Grant Number JPMJSP2125.

\appendix

\section{Loop Functions \texorpdfstring{$B_0$}{B0} and \texorpdfstring{$B_1$}{B1}} 
\label{app:res_B0_B1}
Here, we show the results of $B_0(m_A,m_B,m_S)$ and $B_1(m_A,m_B,m_S)$ expressed as combinations of the three-loop vacuum integrals, $J[n_1,n_2,n_3,n_4,n_5,n_6]$, in Eq.~(\ref{eq:vacuum_int}).
$B_0(m_A,m_B,m_S)$ and $B_1(m_A,m_B,m_S)$ are given as follows:
\begin{align}
    & \hspace{-0.5cm}B_0(m_A,m_B,m_S)
    \nonumber
    \\
    =&
    \frac{8(d-3)}{d+2} 
    (
    J[2,2,3,1,-1,1]
    -
    2 J[2,1,3,1,0,1]
    )
    +
    \frac{8(d^2+d-12)}{d(d+2)} J[1,1,3,1,0,2] 
    \nonumber
    \\   
    &+
    \frac{8(d-8)}{d(d+2)}
    (
     J[2,1,2,1,1,1]
     +
     J[2,2,2,1,0,1]
     -
     J[1,2,2,1,1,1]
     -
     J[2,1,2,1,0,2]
    )
    \nonumber
    \\
    &-
    \frac{8(d-4)}{d} (J[1,1,3,1,1,1] + J[2,2,3,1,0,0])
     -
    \frac{8(d+12)}{d(d+2)} J[1,2,3,1,0,1]
    \nonumber
    \\
    &+
    \frac{16(d-8)}{d(d+2)} (J[3,1,2,1,0,1] - J[3,1,3,1,0,0])
     -
    \frac{16}{d} J[2,0,3,1,1,1]
    \nonumber
    \\
    &+
    \frac{8(2 d^2 - 3d -4)}{d(d+2)} J[2,1,3,1,1,0]
    +
    (n_1 \leftrightarrow n_2,\; n_5 \leftrightarrow n_6)\,,
    \label{App:B0}
\end{align}
and
\begin{align}
    &\hspace{-0.5cm}B_1(m_A,m_B,m_S)
    \nonumber
    \\
    =&
    \frac{16(d-3)}{d+2} 
    (
    J[2,2,4,1,-1,1]
    -
    2 J[2,1,4,1,0,1]
    )
    +
    \frac{16(d^2+d-12)}{d(d+2)} J[1,1,4,1,0,2] 
    \nonumber
    \\    
    &+
    \frac{16(d-8)}{d(d+2)}
    (
     J[2,1,3,1,1,1]
     +
     J[2,2,3,1,0,1]
     -
     J[1,2,3,1,1,1]
     -
     J[2,1,3,1,0,2]
    )
    \nonumber
    \\
    &-
    \frac{16(d-4)}{d} (J[1,1,4,1,1,1] + J[2,2,4,1,0,0])
     -
    \frac{16(d+12)}{d(d+2)} J[1,2,4,1,0,1]
    \nonumber
    \\
    &+
    \frac{32(d-8)}{d(d+2)} (J[3,1,3,1,0,1] - J[3,1,4,1,0,0])
     -
    \frac{32}{d} J[2,0,4,1,1,1]
    \nonumber
    \\
    &+
    \frac{16(2 d^2 - 3d -4)}{d(d+2)} J[2,1,4,1,1,0]
    +
    (n_1 \leftrightarrow n_2,\; n_5 \leftrightarrow n_6)\,,
    \label{App:B1}
\end{align}
where the terms of $(n_1 \leftrightarrow n_2,\; n_5 \leftrightarrow n_6)$ represent the contributions obtained by exchanging $\psi_A$ and $\psi_B$. 
While some of $J[n_1,n_2,n_3,n_4,n_5,n_6]$ are divergent in the limit of $d\rightarrow 4$, 
$B_0(m_A,m_B,m_S)$ and $B_1(m_A,m_B,m_S)$ are finite.

\bibliographystyle{utphys28mod}
\bibliography{ref}

@article{Banno:2024apv,
    author = "Banno, Tatsuya and Hisano, Junji and Kitahara, Teppei and Ogawa, Kiyoto and Osamura, Naohiro",
    title = "{Impact of the electroweak Weinberg operator on the electric dipole moment of electron}",
    eprint = "2408.02375",
    archivePrefix = "arXiv",
    primaryClass = "hep-ph",
    reportNumber = "CHIBA-EP-266, IPMU24-0034",
    doi = "10.1007/JHEP02(2025)082",
    journal = "JHEP",
    volume = "02",
    pages = "082",
    year = "2025"
}

@article{Dekens:2019ept,
    author = "Dekens, Wouter and Stoffer, Peter",
    title = "{Low-energy effective field theory below the electroweak scale: matching at one loop}",
    eprint = "1908.05295",
    archivePrefix = "arXiv",
    primaryClass = "hep-ph",
    doi = "10.1007/JHEP10(2019)197",
    journal = "JHEP",
    volume = "10",
    pages = "197",
    year = "2019",
    note = "[Erratum: JHEP 11, 148 (2022)]"
}

@article{Abe:2024mwa,
    author = "Abe, Tomohiro and Sato, Ryosuke and Yamanaka, Takumu",
    title = "{Composite dark matter with forbidden annihilation}",
    eprint = "2404.03963",
    archivePrefix = "arXiv",
    primaryClass = "hep-ph",
    reportNumber = "OU-HET-1219",
    doi = "10.1007/JHEP09(2024)064",
    journal = "JHEP",
    volume = "09",
    pages = "064",
    year = "2024"
}

@article{Cirelli:2009uv,
    author = "Cirelli, Marco and Strumia, Alessandro",
    title = "{Minimal Dark Matter: Model and results}",
    eprint = "0903.3381",
    archivePrefix = "arXiv",
    primaryClass = "hep-ph",
    reportNumber = "IFUP-TH-2009-04, SACLAY-T09-010",
    doi = "10.1088/1367-2630/11/10/105005",
    journal = "New J. Phys.",
    volume = "11",
    pages = "105005",
    year = "2009"
}

@article{Abe:2017sam,
    author = "Abe, Tomohiro and Hisano, Junji and Nagai, Ryo",
    title = "{Model independent evaluation of the Wilson coefficient of the Weinberg operator in QCD}",
    eprint = "1712.09503",
    archivePrefix = "arXiv",
    primaryClass = "hep-ph",
    reportNumber = "IPMU17-0184, TU-1055",
    doi = "10.1007/JHEP03(2018)175",
    journal = "JHEP",
    volume = "03",
    pages = "175",
    year = "2018",
    note = "[Erratum: JHEP 09, 020 (2018)]"
}

@article{tHooft:1972tcz,
    author = "'t Hooft, Gerard and Veltman, M. J. G.",
    title = "{Regularization and Renormalization of Gauge Fields}",
    doi = "10.1016/0550-3213(72)90279-9",
    journal = "Nucl. Phys. B",
    volume = "44",
    pages = "189--213",
    year = "1972"
}

@article{Hiramoto:2022fyg,
    author = "Hiramoto, A. and others",
    title = "{SiPM module for the ACME III electron EDM search}",
    eprint = "2210.05727",
    archivePrefix = "arXiv",
    primaryClass = "physics.ins-det",
    doi = "10.1016/j.nima.2022.167513",
    journal = "Nucl. Instrum. Meth. A",
    volume = "1045",
    pages = "167513",
    year = "2023"
}

@article{Martin:2001vx,
    author = "Martin, Stephen P.",
    title = "{Two Loop Effective Potential for a General Renormalizable Theory and Softly Broken Supersymmetry}",
    eprint = "hep-ph/0111209",
    archivePrefix = "arXiv",
    reportNumber = "FERMILAB-PUB-01-348-T",
    doi = "10.1103/PhysRevD.65.116003",
    journal = "Phys. Rev. D",
    volume = "65",
    pages = "116003",
    year = "2002"
}

@article{Ford:1992pn,
    author = "Ford, C. and Jack, I. and Jones, D. R. T.",
    title = "{The Standard model effective potential at two loops}",
    eprint = "hep-ph/0111190",
    archivePrefix = "arXiv",
    reportNumber = "NSF-ITP-92-21, LTH-281",
    doi = "10.1016/0550-3213(92)90165-8",
    journal = "Nucl. Phys. B",
    volume = "387",
    pages = "373--390",
    year = "1992",
    note = "[Erratum: Nucl.Phys.B 504, 551--552 (1997)]"
}

@article{Espinosa:2000df,
    author = "Espinosa, Jose Ramon and Zhang, Ren-Jie",
    title = "{Complete two loop dominant corrections to the mass of the lightest CP even Higgs boson in the minimal supersymmetric standard model}",
    eprint = "hep-ph/0003246",
    archivePrefix = "arXiv",
    reportNumber = "IFT-UAM-CSIC-00-09, MADPH-00-1158",
    doi = "10.1016/S0550-3213(00)00421-1",
    journal = "Nucl. Phys. B",
    volume = "586",
    pages = "3--38",
    year = "2000"
}

@article{Jenkins:2017dyc,
    author = "Jenkins, Elizabeth E. and Manohar, Aneesh V. and Stoffer, Peter",
    title = "{Low-Energy Effective Field Theory below the Electroweak Scale: Anomalous Dimensions}",
    eprint = "1711.05270",
    archivePrefix = "arXiv",
    primaryClass = "hep-ph",
    doi = "10.1007/JHEP01(2018)084",
    journal = "JHEP",
    volume = "01",
    pages = "084",
    year = "2018"
}

@article{ACME:2018yjb,
    author = "Andreev, V. and others",
    collaboration = "ACME",
    title = "{Improved limit on the electric dipole moment of the electron}",
    doi = "10.1038/s41586-018-0599-8",
    journal = "Nature",
    volume = "562",
    number = "7727",
    pages = "355--360",
    year = "2018"
}

@article{Roussy:2022cmp,
    author = "Roussy, Tanya S. and others",
    title = "{An improved bound on the electron\textquoteright{}s electric dipole moment}",
    eprint = "2212.11841",
    archivePrefix = "arXiv",
    primaryClass = "physics.atom-ph",
    doi = "10.1126/science.adg4084",
    journal = "Science",
    volume = "381",
    number = "6653",
    pages = "adg4084",
    year = "2023"
}

@article{Barr:1990vd,
    author = "Barr, Stephen M. and Zee, A.",
    title = "{Electric Dipole Moment of the Electron and of the Neutron}",
    reportNumber = "NSF-ITP-90-46",
    doi = "10.1103/PhysRevLett.65.21",
    journal = "Phys. Rev. Lett.",
    volume = "65",
    pages = "21--24",
    year = "1990",
    note = "[Erratum: Phys.Rev.Lett. 65, 2920 (1990)]"
}

@article{Nagata:2014wma,
    author = "Nagata, Natsumi and Shirai, Satoshi",
    title = "{Higgsino Dark Matter in High-Scale Supersymmetry}",
    eprint = "1410.4549",
    archivePrefix = "arXiv",
    primaryClass = "hep-ph",
    reportNumber = "DESY-14-180, FTPI-MINN-14-37, IPMU14-0320",
    doi = "10.1007/JHEP01(2015)029",
    journal = "JHEP",
    volume = "01",
    pages = "029",
    year = "2015"
}

@article{Ema:2022yra,
    author = "Ema, Yohei and Gao, Ting and Pospelov, Maxim",
    title = "{Standard Model Prediction for Paramagnetic Electric Dipole Moments}",
    eprint = "2202.10524",
    archivePrefix = "arXiv",
    primaryClass = "hep-ph",
    reportNumber = "UMN-TH-4115/22, FTPI-MINN-22-06",
    doi = "10.1103/PhysRevLett.129.231801",
    journal = "Phys. Rev. Lett.",
    volume = "129",
    number = "23",
    pages = "231801",
    year = "2022"
}

@inproceedings{Alarcon:2022ero,
    author = "Alarcon, Ricardo and others",
    title = "{Electric dipole moments and the search for new physics}",
    booktitle = "{Snowmass 2021}",
    eprint = "2203.08103",
    archivePrefix = "arXiv",
    primaryClass = "hep-ph",
    month = "3",
    year = "2022"
}

@article{Hisano:2014kua,
    author = "Hisano, Junji and Kobayashi, Daiki and Mori, Naoya and Senaha, Eibun",
    title = "{Effective Interaction of Electroweak-Interacting Dark Matter with Higgs Boson and Its Phenomenology}",
    eprint = "1410.3569",
    archivePrefix = "arXiv",
    primaryClass = "hep-ph",
    reportNumber = "IPMU-14-0309",
    doi = "10.1016/j.physletb.2015.01.012",
    journal = "Phys. Lett. B",
    volume = "742",
    pages = "80--85",
    year = "2015"
}

@article{Hisano:2006nn,
    author = "Hisano, Junji and Matsumoto, Shigeki and Nagai, Minoru and Saito, Osamu and Senami, Masato",
    title = "{Non-perturbative effect on thermal relic abundance of dark matter}",
    eprint = "hep-ph/0610249",
    archivePrefix = "arXiv",
    reportNumber = "KEK-TH-1111",
    doi = "10.1016/j.physletb.2007.01.012",
    journal = "Phys. Lett. B",
    volume = "646",
    pages = "34--38",
    year = "2007"
}

@article{Atwood:1990cm,
    author = "Atwood, D. and Burgess, C. P. and Hamazaou, C. and Irwin, B. and Robinson, J. A.",
    title = "{One loop P and T odd W+- electromagnetic moments}",
    reportNumber = "PRINT-90-0358 (BNL), MCGILL-90-16, BNL-44957",
    doi = "10.1103/PhysRevD.42.3770",
    journal = "Phys. Rev. D",
    volume = "42",
    pages = "3770--3777",
    year = "1990"
}

@article{Hoogeveen:1987jn,
    author = "Hoogeveen, F.",
    title = "{A bound on CP violating couplings of the W boson}",
    reportNumber = "MPI-PAE/PTh-25/87",
    month = "3",
    year = "1987",
    note = "MPI-PAE/PTh-25/87, 1987"
}

@article{Boudjema:1990dv,
    author = "Boudjema, F. and Hagiwara, Kaoru and Hamzaoui, C. and Numata, K.",
    title = "{Anomalous moments of quarks and leptons from nonstandard W W gamma couplings}",
    reportNumber = "KEK-TH-254, UDEM-LPN-TH-33, UT-KOMABA-90-18, KEK-PREPRINT-90-39",
    doi = "10.1103/PhysRevD.43.2223",
    journal = "Phys. Rev. D",
    volume = "43",
    pages = "2223--2232",
    year = "1991"
}

@article{Novales-Sanchez:2007rsw,
    author = "Novales-Sanchez, H. and Toscano, J. J.",
    title = "{Effective Lagrangian approach to fermion electric dipole moments induced by a CP-violating WW gamma vertex}",
    eprint = "0712.2008",
    archivePrefix = "arXiv",
    primaryClass = "hep-ph",
    doi = "10.1103/PhysRevD.77.015011",
    journal = "Phys. Rev. D",
    volume = "77",
    pages = "015011",
    year = "2008"
}

@article{DeRujula:1990db,
    author = "De Rujula, A. and Gavela, M. B. and Pene, O. and Vegas, F. J.",
    title = "{Signets of CP violation}",
    reportNumber = "CERN-TH-5908-90, LPTHE-ORSAY-90-41",
    doi = "10.1016/0550-3213(91)90472-A",
    journal = "Nucl. Phys. B",
    volume = "357",
    pages = "311--356",
    year = "1991"
}

@article{Cirelli:2005uq,
    author = "Cirelli, Marco and Fornengo, Nicolao and Strumia, Alessandro",
    title = "{Minimal dark matter}",
    eprint = "hep-ph/0512090",
    archivePrefix = "arXiv",
    reportNumber = "DFTT40-2005, IFUP-TH-2005-34",
    doi = "10.1016/j.nuclphysb.2006.07.012",
    journal = "Nucl. Phys. B",
    volume = "753",
    pages = "178--194",
    year = "2006"
}

@article{Cirelli:2007xd,
    author = "Cirelli, Marco and Strumia, Alessandro and Tamburini, Matteo",
    title = "{Cosmology and Astrophysics of Minimal Dark Matter}",
    eprint = "0706.4071",
    archivePrefix = "arXiv",
    primaryClass = "hep-ph",
    reportNumber = "IFUP-TH-2007-12, SACLAY-T07-052",
    doi = "10.1016/j.nuclphysb.2007.07.023",
    journal = "Nucl. Phys. B",
    volume = "787",
    pages = "152--175",
    year = "2007"
}

@article{PDG:2024,
    author = "Navas, S. and others",
    collaboration = "Particle Data Group",
    title = "{Review of particle physics}",
    doi = "10.1103/PhysRevD.110.030001",
    journal = "Phys. Rev. D",
    volume = "110",
    number = "3",
    pages = "030001",
    year = "2024"
}

@article{Banno:2023yrd,
    author = "Banno, Tatsuya and Hisano, Junji and Kitahara, Teppei and Osamura, Naohiro",
    title = "{Closer look at the matching condition for radiative QCD \ensuremath{\theta} parameter}",
    eprint = "2311.07817",
    archivePrefix = "arXiv",
    primaryClass = "hep-ph",
    reportNumber = "IPMU23-0042",
    doi = "10.1007/JHEP02(2024)195",
    journal = "JHEP",
    volume = "02",
    pages = "195",
    year = "2024"
}

@article{Breitenlohner:1977hr,
    author = "Breitenlohner, P. and Maison, D.",
    title = "{Dimensional Renormalization and the Action Principle}",
    doi = "10.1007/BF01609069",
    journal = "Commun. Math. Phys.",
    volume = "52",
    pages = "11--38",
    year = "1977"
}

@article{ATLAS:2022rme,
    author = "Aad, Georges and others",
    collaboration = "ATLAS",
    title = "{Search for long-lived charginos based on a disappearing-track signature using 136 fb$^{-1}$ of pp collisions at $\sqrt{s}$~=~13~TeV with the ATLAS detector}",
    eprint = "2201.02472",
    archivePrefix = "arXiv",
    primaryClass = "hep-ex",
    reportNumber = "CERN-EP-2021-209",
    doi = "10.1140/epjc/s10052-022-10489-5",
    journal = "Eur. Phys. J. C",
    volume = "82",
    number = "7",
    pages = "606",
    year = "2022"
}

@article{Maierhofer:2017gsa,
    author = {Maierh{\"o}fer, Philipp and Usovitsch, Johann and Uwer, Peter},
    title = "{Kira{\textemdash}A Feynman integral reduction program}",
    eprint = "1705.05610",
    archivePrefix = "arXiv",
    primaryClass = "hep-ph",
    doi = "10.1016/j.cpc.2018.04.012",
    journal = "Comput. Phys. Commun.",
    volume = "230",
    pages = "99--112",
    year = "2018"
}

@article{Klappert:2020nbg,
    author = {Klappert, Jonas and Lange, Fabian and Maierh{\"o}fer, Philipp and Usovitsch, Johann},
    title = "{Integral reduction with Kira 2.0 and finite field methods}",
    eprint = "2008.06494",
    archivePrefix = "arXiv",
    primaryClass = "hep-ph",
    reportNumber = "TTK-20-24, P3H-20-041, FR-PHENO-2020-11, MITP/20-044",
    doi = "10.1016/j.cpc.2021.108024",
    journal = "Comput. Phys. Commun.",
    volume = "266",
    pages = "108024",
    year = "2021"
}

@article{Lange:2025fba,
    author = "Lange, Fabian and Usovitsch, Johann and Wu, Zihao",
    title = "{Kira 3: integral reduction with efficient seeding and optimized equation selection}",
    eprint = "2505.20197",
    archivePrefix = "arXiv",
    primaryClass = "hep-ph",
    reportNumber = "ZU-TH 39/25, HU-EP-25/17-RTG",
    month = "5",
    year = "2025"
}

@incollection{Fermat,
    author = "Lewis, Robert H.",
    title = "{Fermat, A Computer Algebra System for Polynomial and Matrix Computation}",
    note = "{URL: \url{http://home.bway.net/lewis/}}"
}

@article{Martin:2016bgz,
    author = "Martin, Stephen P. and Robertson, David G.",
    title = "{Evaluation of the general 3-loop vacuum Feynman integral}",
    eprint = "1610.07720",
    archivePrefix = "arXiv",
    primaryClass = "hep-ph",
    doi = "10.1103/PhysRevD.95.016008",
    journal = "Phys. Rev. D",
    volume = "95",
    number = "1",
    pages = "016008",
    year = "2017"
}

@article{Jenkins:2017jig,
    author = "Jenkins, Elizabeth E. and Manohar, Aneesh V. and Stoffer, Peter",
    title = "{Low-Energy Effective Field Theory below the Electroweak Scale: Operators and Matching}",
    eprint = "1709.04486",
    archivePrefix = "arXiv",
    primaryClass = "hep-ph",
    doi = "10.1007/JHEP03(2018)016",
    journal = "JHEP",
    volume = "03",
    pages = "016",
    year = "2018",
    note = "[Erratum: JHEP 12, 043 (2023)]"
}

@article{Pospelov:2013sca,
    author = "Pospelov, Maxim and Ritz, Adam",
    title = "{CKM benchmarks for electron electric dipole moment experiments}",
    eprint = "1311.5537",
    archivePrefix = "arXiv",
    primaryClass = "hep-ph",
    doi = "10.1103/PhysRevD.89.056006",
    journal = "Phys. Rev. D",
    volume = "89",
    number = "5",
    pages = "056006",
    year = "2014"
}

@article{Banno:2025pfq,
    author = "Banno, Tatsuya and Hisano, Junji and Kitahara, Teppei and Ogawa, Kiyoto and Osamura, Naohiro",
    title = "{Two-loop corrections to QCD {\ensuremath{\theta}} angle from evanescent operator in the BMHV scheme}",
    eprint = "2502.14500",
    archivePrefix = "arXiv",
    primaryClass = "hep-ph",
    reportNumber = "CHIBA-EP-267, IPMU24-0005",
    doi = "10.1007/JHEP09(2025)135",
    journal = "JHEP",
    volume = "09",
    pages = "135",
    year = "2025"
}

\end{document}